\documentclass[sigconf]{acmart}

\AtBeginDocument{%
  \providecommand\BibTeX{{%
    \normalfont B\kern-0.5em{\scshape i\kern-0.25em b}\kern-0.8em\TeX}}}

\setcopyright{acmcopyright}
\copyrightyear{2021}
\acmYear{2021}
\acmDOI{10.1145/1122445.1122456}

\copyrightyear{2021}
\acmYear{2021}
\setcopyright{acmcopyright}
\acmConference[CHI '21]{CHI Conference on Human Factors in Computing Systems}{May 8--13, 2021}{Yokohama, Japan}
\acmBooktitle{CHI Conference on Human Factors in Computing Systems (CHI '21), May 8--13, 2021, Yokohama, Japan}
\acmPrice{15.00}
\acmDOI{10.1145/3411764.3445448}
\acmISBN{978-1-4503-8096-6/21/05}

\begin{document}


\title[Heterogeneous Stroke]{Heterogeneous Stroke: Using Unique Vibration Cues to Improve the Wrist-Worn Spatiotemporal Tactile Display}

\author{Taejun Kim}
\affiliation{%
  \institution{HCI Lab, KAIST}
  \city{Daejeon}
  \country{South Korea}}
\email{taejun.kim@kaist.ac.kr}

\author{Youngbo Aram Shim}
\affiliation{%
  \institution{HCI Lab, KAIST}
  \city{Daejeon}
  \country{South Korea}
}
\email{youngbo.shim@kaist.ac.kr}

\author{Geehyuk Lee}
\affiliation{%
  \institution{HCI Lab, KAIST}
  \city{Daejeon}
  \country{South Korea}
}
\email{geehyuk@gmail.com}

\renewcommand{\shortauthors}{Kim et al.}

\begin{abstract}

Beyond a simple notification of incoming calls or messages, more complex information such as alphabets and digits can be delivered through spatiotemporal tactile patterns (STPs) on a wrist-worn tactile display (WTD) with multiple tactors. However, owing to the limited skin area and spatial acuity of the wrist, frequent confusions occur between closely located tactors, resulting in a low recognition accuracy. Furthermore, the accuracies reported in previous studies have mostly been measured for a specific posture and could further decrease with free arm postures in real life. Herein, we present Heterogeneous Stroke, a design concept for improving the  recognition accuracy of STPs on a WTD. By assigning unique vibrotactile stimuli to each tactor, the confusion between tactors can be reduced. Through our implementation of Heterogeneous Stroke, the alphanumeric characters could be delivered with high accuracy (93.8\% for 26 alphabets and 92.4\% for 10 digits) across different arm postures.

\end{abstract}



\maketitle

\section{Introduction}

As wrist-worn devices, e.g., smartwatches, have become popularized, various tactile communication on the wrist-worn tactile display (WTD) has been studied \cite{lee2015investigating, shim2019using, liao2016edgevib, lee2010buzzwear}. A typical WTD in the market uses the entire frame of the device as a single vibrotactor and mainly employs the temporal profile, i.e., rhythm, of the vibration to design distinguishable tactile patterns. However, as the number of patterns increases, the temporal patterns become lengthy and difficult to interpret and memorize.

For a more effective communication, researchers have proposed spatiotemporal tactile patterns (STPs) by arranging multiple tactors in a triangular \cite{lee2010buzzwear}, square \cite{liao2016edgevib, shim2018exploring}, or grid \cite{lee2015investigating, shim2019using, lee2009mobile} layout on the wrist. As a consequence, the information transmission efficiency could be increased, and more intuitive communication becomes possible by designing the spatial form of STP to be aligned with the actual meaning. For instance, the directional information for navigation was delivered by the movement of stimulus \cite{lee2015investigating, shim2019using, lee2009mobile, ion2015skin}, and the alphanumeric information was delivered with STPs whose ``spatiotemporal stroke'' is close to the shape of the corresponding character \cite{liao2016edgevib, lee2009mobile, ion2015skin}.

The recognition accuracies were acceptable for relatively simple patterns, i.e., 92 \% upon discrimination of four directional patterns in the up, down, left, and right directions \cite{lee2015investigating}, and 95 \% upon discrimination of 24 patterns using levels of four parameters including the intensity and moving direction \cite{lee2010buzzwear}. However, for complex patterns designed based on their 2D-shape like alphabet patterns, the accuracy was too low (71 \% upon discrimination of 26 EdgeWrite \cite{wobbrock2003edgewrite} alphabet patterns \cite{liao2016edgevib}) to be used in real life.

The main causes of the poor recognition accuracy are the limited skin area and spatial acuity of the wrist. Chen et al. \cite{chen2008tactor} showed that out of nine vibrotactors arranged at 25 mm intervals, only two could be reliably localized on the dorsal and volar wrist. Owing to a narrow skin area, tactors are closely located, resulting in frequent confusion. Furthermore, most of the previous studies measured the accuracy in a specific arm posture \cite{lee2015investigating, shim2019using, liao2016edgevib, ion2015skin, shim2018exploring}. However, existing studies have shown that tactile perception can be influenced by the posture \cite{lawson2014you, scocchia2009influence, cody2010tactile} or movement \cite{chen2018effect, pakkanen2008perception, post1994perception} of the body parts. Considering that users can take free postures in real life, the accuracy reported in a specific posture can be further decreased. We can state that the interaction is ready to be used in real life when usable accuracies are ensured across various postures.

In this paper, we propose Heterogeneous Stroke, a design concept for improving the recognition accuracy of STPs on a WTD. Heterogeneous Stroke assigns unique vibrotactile stimuli to each tactor and utilizes them to effectively recognize STPs. By making the stimulus of each tactor more distinguishable, confusions between tactors can be reduced. To implement the proposed concept, we designed four unique vibrotactile stimuli by combining two levels of frequency (170 and 300Hz) and two levels of roughness (with and without roughness) through an amplitude modulation of the waveform. We then experimentally studied the effect of Heterogeneous Stroke in terms of accuracy improvement.

We first conducted a preliminary study to investigate the effect of the arm posture on the recognition accuracy of the STPs to derive requirements for a usable WTD. Through the task of recognizing the EdgeWrite \cite{wobbrock2003edgewrite} pattern set, which can intuitively deliver alphanumeric characters on a 2 x 2 tactor array, we observed that the accuracy can be significantly affected by the arm posture of the user. In User Study 1, we investigated the effect of Heterogeneous Stroke in terms of an accuracy improvement across different arm postures. The three-point-stroke set, which consists of every possible pattern that stimulates three consecutive points in a 2 x 2 tactor array, was used for the task because it forms the elements of any STP having a 2D-shape. Finally, in User Study 2, using the EdgeWrite pattern set again, we confirmed that 26 alphabets and 10 digits can be delivered with high accuracy (93.8 \% for alphabets and 92.4 \% for digits) through our implementation of Heterogeneous Stroke. The main contributions of this study are as follows:
\begin{itemize}
  \item We proposed a design concept of Heterogeneous Stroke that utilizes unique vibration cues to improve the recognition accuracy of STPs on a WTD.
  \item Through our implementation of Heterogeneous Stroke, we empirically showed that the proposed design significantly improves the recognition accuracy of STPs on a WTD.
\end{itemize}

\section{Related Work}

We first review various STPs proposed within the wearable tactile display domain. We then review previous studies that imply that the user's posture can influence a tactile pattern recognition. Lastly, we review previous work that facilitates different vibration parameters for the information encoding and position the current study among them.

\subsection{Spatiotemporal Tactile Patterns on Wearable Tactile Displays}

Researchers have proposed various STP designs by locating multiple tactors on the wearable tactile display for a more effective information transmission.

The previous STP designs can be divided into two different purposes. One purpose is to explore the combinations of spatiotemporal parameters to produce a large number of usable patterns within a given tactor layout. Shim et al. \cite{shim2018exploring} and Lee et al. \cite{lee2010buzzwear} designed 16 and 24 patterns, respectively, by combining the levels of parameters including the starting point and moving direction of the STPs. More studies \cite{matscheko2010tactor, park2018haptic} have designed a pattern set in a similar way and reported the information transfer they achieved. These studies aimed to explore the information transmission capacity of the designed wearable tactile displays, rather than support a specific application.

In other approaches, researchers exploited STPs for intuitive information encoding under a target scenario. The spatial form of the patterns was designed to be aligned with the actual meaning. Various studies \cite{paneels2013what, lee2015investigating, lee2009mobile, tang2020design, jones2009vibrotactile, shim2018exploring, shim2019using} have delivered directional information for navigation based on the movement of stimuli. Information such as alphabets \cite{lee2009mobile, liao2016edgevib, ion2015skin}, numbers \cite{liao2016edgevib}, and phonemes \cite{turcott2018efficient, zhao2018coding} has also been delivered with STPs whose spatial form implies the shape of the letter or position of the tongue when pronounced. In particular, Liao et al. \cite{liao2016edgevib} attempted to deliver all 26 alphabets and 10 digits on the wrist by employing EdgeWrite pattern, which is a set of uni-stroke paths that connect the points of a 2 $\times$ 2 array. These designs consider the types of information to be conveyed at the application level and aim for users to intuitively understand and learn patterns through shape recognition. 

In the latter designs, the recognition accuracy was reported to be acceptable for a relatively simple pattern set, e.g., 92 \% upon the discrimination of four directional patterns \cite{lee2015investigating}. However, for complex patterns that are based on their 2D-shape, achieving a usable recognition accuracy was difficult or required additional resources. In the study of Liao et al. \cite{liao2016edgevib}, the average duration of the pattern was increased from 2.2 to 2.9 s to improve the low discrimination accuracy of 70.7 \% for the 26 EdgeWrite alphabet patterns up to 85.9 \%. 

Our proposed solution, Heterogeneous Stroke, achieves a high recognition accuracy with complex STP set such as those used in the EdgeVib \cite{liao2016edgevib} study, even without lengthening the duration of the patterns.

\subsection{Postures and Tactile Pattern Recognition}

Applying wearable vibrotactile haptics technology into the wild requires careful design considerations. Environmental differences between the lab condition and real world have been described as one of the factors that makes the experimental results from the lab unreliable \cite{blum2019getting}. In this sense, the body posture of the user may affect the tactile recognition.

Researchers have observed the influence of the user's physical activity, e.g., running \cite{chen2018effect} and biking \cite{pakkanen2008perception}, on the tactile pattern recognition. In particular, Post et al. \cite{post1994perception} showed that the tactile ability to detect and scale the vibration stimuli can deteriorate when people repeatedly flex and extend the elbow with their arms. The motor task that affected the tactile perception could be interpreted as a repetition of switch between two body postures. Cody et al. \cite{cody2010tactile} reported that the skin stretch of the wrist caused by bending the hand can reduce the tactile spatial acuity, implying that even a static posture can influence the tactile recognition ability.

In addition to the physical influence, the body posture may also affect the neural process of decoding the tactile stimuli. The localization of tactile stimuli requires multiple information to be integrated \cite{medina2010maps}. Systematic distortions can occur in localization tasks \cite{longo2015implicit, mancini2011supramodal}, and various factors can influence this process \cite{medina2010maps}. Previous studies have shown that the head orientation induces a systematic bias of tactile localization in the waist \cite{ho2007head} and forearm \cite{pritchett2011perceived}. Lawson et al. \cite{lawson2014you} and Schocchia et al. \cite{scocchia2009influence} also showed that the head orientation with a certain arm posture could influence the response time and error rates of a haptic identification task. This tendency of interpreting the haptic signals based on the head-centered reference frame may alter the perception of identical tactile patterns owing to the body postures, or in other words the spatial relationship between the head and the arm.

Diverse postures can be applied in our daily life. One can put an arm down to hold a shopping bag, raise it up to grab a bus handle, reach it forward to grasp a steering wheel, or simply bring it in front of your body to check the time on a watch. However, previous WTD studies have mostly evaluated the recognition accuracy in a specific arm posture. Participants were guided to put their arm forward on a desk while sitting \cite{lee2015investigating, shim2018exploring, shim2019using, ion2015skin}, put it close to their body as if looking at a wristwatch \cite{liao2016edgevib}, or simply rest it with a comfortable posture without much control \cite{lee2010buzzwear, matscheko2010tactor}. In the preliminary study, we first investigated whether the arm posture affects the pattern recognition accuracy of the STPs. Three arm postures used in previous studies, as shown in Figure \ref{Armpostures}, were tested.

\subsection{Using Different Vibrotactile Parameters for Wearable Tactile Display}

Researchers have explored various tactile attributes to expand the information transmission capacity of the tactile display \cite{tan2020methodology}. The physical parameters such as frequency \cite{brewster2004tactons, brown2006multidimensional, tan1999information}, duration \cite{brewster2004tactons}, waveform \cite{brewster2004tactons}, amplitude \cite{brewster2004tactons, lee2010buzzwear, tan1999information}, and spatial location \cite{brown2006multidimensional} and higher-level parameters such as roughness \cite{brewster2004tactons, brown2006multidimensional} and rhythm \cite{brewster2004tactons, brown2006multidimensional, lee2010buzzwear} using a signal modulation were explored for the vibration.

Reed et al. \cite{reed2018phonemic} and Tan et al. \cite{tan2020acquisition} designed a forearm-mounted device consisting of 4 $\times$ 6 vibrotactor array, and effectively conveyed 39 English phonemes with combinations of parameters such as frequency and duration. Luzhnica et al. \cite{luzhnica2019optimising} designed a glove-type wearable display with several vibrotactors attached, and conveyed 26 alphabets through the back of the hand with high accuracy using combinations of spatial locations that were stimulated. Within the WTD domain, BuzzWear \cite{lee2010buzzwear} designed 24 patterns with combinations of parameters such as the amplitude and rhythm in a triangular tactor layout. These patterns can be conveyed with a high accuracy with a pattern duration of ~2.3s. 

The idea of Heterogeneous Stroke is different from that of previous approaches \cite{reed2018phonemic, tan2020acquisition, luzhnica2019optimising, lee2010buzzwear, shim2018exploring} in that it focuses on solving the positional ambiguity of tactors by assigning a unique vibration to each tactor. Previous studies explored exhaustive combinations of vibration parameters to increase the number of usable patterns (e.g., an ``i5'' tactor in Reed et al.'s study \cite{reed2018phonemic} can present 6 types of vibration), where as Heterogeneous Stroke is a redundant approach that can be applied to the existing pattern set to strengthen the distinction between tactor positions. Heterogeneous Stroke is an effective approach to overcome the limitation of the wrist space. It helps users better recognize a ``spatiotemporal stroke'' by reducing the positional ambiguity of the tactors. 

\section{Heterogeneous Stroke}

\begin{figure}[b]
  \centering
  \includegraphics[width=8cm]{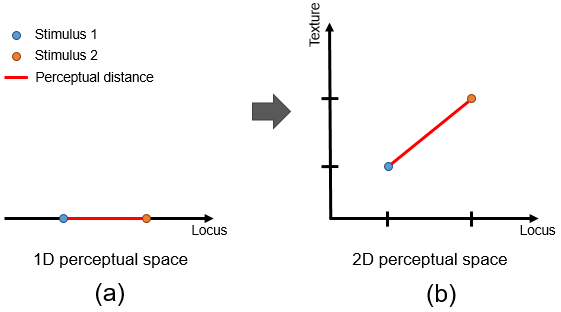}
  \caption{Extending the perceptual space from 1D to 2D.}
  \label{Perceptualspace}
  
\end{figure}

The wrist, a representative body part where wearable devices are worn, has a potential as a space for tactile communication but at the same time has a limitation of a small skin area and low spatial acuity. Owing to such limitation, the distance between multiple tactors becomes smaller than the two-point discrimination threshold of the forearm (i.e., 40mm \cite{weinstein1968intensive} using Von Frey filaments), which leads to frequent confusions between the tactors. Among the nine tactors arranged on the wrist at 25mm intervals, only two could be reliably localized in Chen et al.'s \cite{chen2008tactor} study.

Heterogeneous Stroke is a design concept that assigns unique vibrotactile stimuli to each tactor and utilizes them to effectively recognize STPs. By making the stimulus of each tactor more distinguishable, the confusions between tactors can be reduced. As shown in Figure \ref{Perceptualspace}a, the perceptual space \cite{hollins2000individual} is in 1D when only the locus of the stimulus is used to distinguish between two stimuli. If we provide another perceptual dimension, e.g., the texture of the stimulus, the perceptual space can be extended to 2D. Because the perceptual distance \cite{hollins2000individual} between two stimuli can expand in a 2D perceptual space, the confusion between the two stimuli are expected to be reduced.

Azadi and Jones \cite{azadi2014evaluating} explored the basic vibrotactile parameters, including frequency, amplitude, waveform, and temporal profile (i.e., rhythm) with a single tactor. Among the nine vibrotactile Tactons \cite{brewster2004tactons} they designed, five or six can be distinguished both on the finger and forearm. However, when multiple tactors are arranged closely on the wrist and produce STPs, the situation becomes more challenging. Initially, a certain level of intensity should be ensured not to cause confusion between tactors again. For instance, when 2 and 5 V are used to control the vibration intensity, 2 V will possibly worsen the ambiguity of the position. Therefore, the variation in amplitude was difficult to utilize. In addition, to apply the variation of rhythm, the burst of vibration in each rhythm should be sufficient in length. However, this can make the entire STP too long.

To implement the Heterogeneous Stroke, we utilized two vibrotactile parameters, frequency and roughness \cite{brown2005first}, to design unique vibrotactile stimuli considering insight from the previous studies \cite{brown2005first, brown2006multidimensional, park2011perceptual}. The vibrotactile roughness can be created using an amplitude modulation of the waveform. Previous research \cite{weisenberger1986sensitivity} reported that the amplitude modulated version of vibration was felt to be ``rougher'' than the un-modulated version. By combining two levels of frequency (170 and 300 Hz) and two levels of roughness (with and without roughness), four unique vibrotactile stimuli were designed to implement the Heterogeneous Stroke.

\subsection{Apparatus}

\begin{figure}[b]
  \centering
  \includegraphics[width=8.5cm]{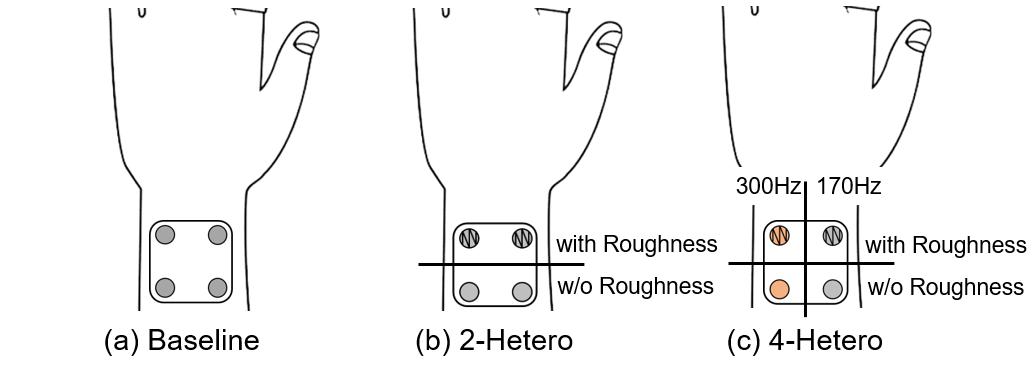}
  \caption{Three types of method used in User Study 1.}
  \label{Threemethods}
  \Description[Three types of method used in User Study 1]{The illustration of three hands and each type of methods with the vibration motors. For example, the Baseline method was described with four same-colored vibration motor icons. }
\end{figure}

The 170 Hz was the resonance frequency of the used motor, and 300 Hz was chosen as a value sufficiently far from 170 Hz within the perceptual range where the stimulus can clearly be felt. To balance the perceived intensity between the two conditions, the vibration motors using 300 and 170 Hz were driven by 9 and 5 V, respectively. The driving voltage was determined through an informal pilot study. To implement the roughness of the vibration, we used a 12.5-Hz on/off modulation of the waveform. Park et al. \cite{park2011perceptual} reported that, at a modulation frequency of 1-20 Hz, the amplitude-modulated vibrotactile stimuli were perceptually dissimilar compared to the non-modulated version. Within this range, we selected a modulation frequency in which the roughness of vibration can be clearly felt.

Considering that the spatial acuity of the longitudinal axis is significantly lower than that of the transverse axis on the wrist \cite{oakley2006determining}, we used a parameter judged to make the stimulus more distinguishable than the other along the longitudinal axis. Through an informal pilot study, the variation of roughness was judged to be more effective than that of frequency. As shown in Figure \ref{Threemethods}c, the two levels of frequency and roughness were used to reduce confusion in the transverse and longitudinal axis of the wrist, respectively.

For a comparative analysis in User Study 1, we set up three methods. The \textit{Baseline} method uses only normal vibration (Figure \ref{Threemethods}a), \textit{2-Hetero} uses two unique vibrations made by two levels of roughness (Figure \ref{Threemethods}b), and \textit{4-Hetero} uses four unique vibrations made by two levels of roughness and frequency (Figure \ref{Threemethods}c).

\section{Preliminary Study: Effect of Posture}

To derive the requirements for a usable WTD, we first investigate the effect of posture on the recognition accuracy of STPs in the preliminary study.

\subsection{Independent Variables}

\begin{figure}[b]
  \centering
  \includegraphics[width=8cm]{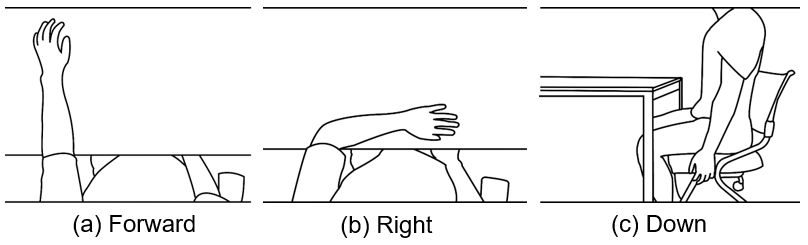}
  \caption{Three arm postures tested in Experiment 1.}
  \label{Armpostures}
  \Description[Three arm postures]{Three arm postures, Forward and Right and Down, are described with a simple illustration.}
\end{figure}

\begin{figure}[t]
  \centering
  \includegraphics[width=8cm]{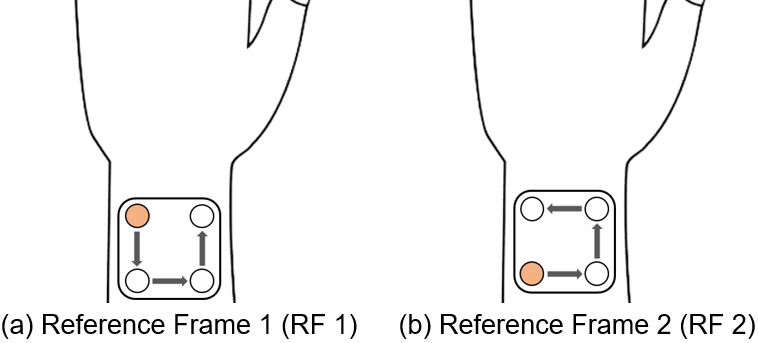}
  \caption{The EdgeWrite pattern for letter U is shown in each type of reference frame: (a) Reference Frame 1 (RF 1) and (b) Reference Frame 2 (RF 2). The pattern starts at a colored point.}
  \label{Referenceframes}
  \Description[Two types of reference frame]{The description of the two types of reference frame with the hand. For reference frame one, the EdgeWrite pattern for letter U is displayed by treating the hand as the top. For reference frame two, it is delivered by treating the direction of twelve o'clock of the wristwatch as the top.}
\end{figure}

We set the arm posture and type of reference frame as independent variables. We tested three arm postures depicted in Figure \ref{Armpostures}, which were mainly used in previous studies. \cite{lee2015investigating, shim2018exploring, shim2019using, ion2015skin, liao2016edgevib, lee2010buzzwear, matscheko2010tactor} 

As depicted in Figure \ref{Referenceframes}, the pattern can be delivered by assuming the side of the watch near the hand as the top (Figure \ref{Referenceframes}a), or interpreting it as an ordinary wristwatch (Figure \ref{Referenceframes}b). Because there is a difference in spatial acuity in the transverse and longitudinal axis of the wrist \cite{oakley2006determining}, we expect that the recognition accuracy of each pattern can be affected by the type of reference frame.

\subsection{Apparatus}

\begin{figure}[t]
  \centering
  \includegraphics[width=7cm]{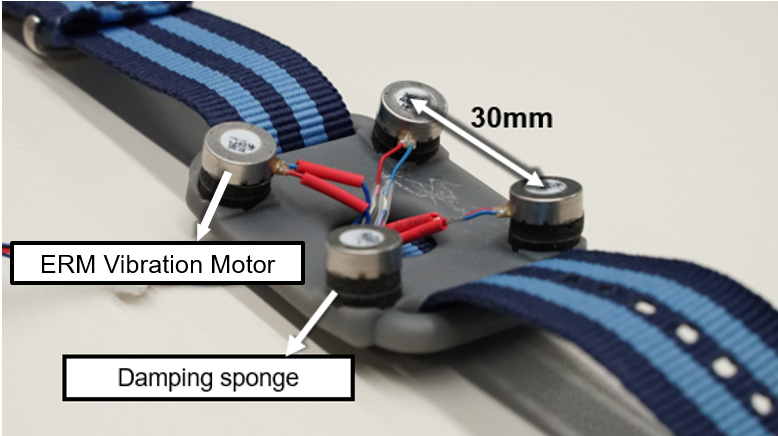}
  \caption{Wrist-worn tactile display prototype.}
  \label{Wtdprototype}
  \Description[The photo of prototype]{The photo of WTD prototype we implemented. Four ERM vibration motors are attached to 3D printed watch frame with the spacing of 30 mm.}
\end{figure}

We implemented a WTD prototype that is identical to that used in Liao et al.'s \cite{liao2016edgevib} study, except that the plastic tips were not attached to each tactor. On a 3D printed watch frame with a size of 40 $\times$ 40 $ mm^2$, four vibration motors of 10 mm in diameter were attached in a 2 $\times$ 2 array form. The distance between the center of the motors was 30 mm, as shown in Figure \ref{Wtdprototype}. Damping sponges were placed between the motors and the frame to isolate the vibration. In the preliminary study, Eccentric Rotating Mass (ERM) vibration motors were used, and each motor was driven with a voltage of 5 V. We used an Arduino UNO as a microcontroller to control the actuators.

\subsection{Tactile Pattern Set}

\begin{figure}[b]
  \centering
  \includegraphics[width=8cm]{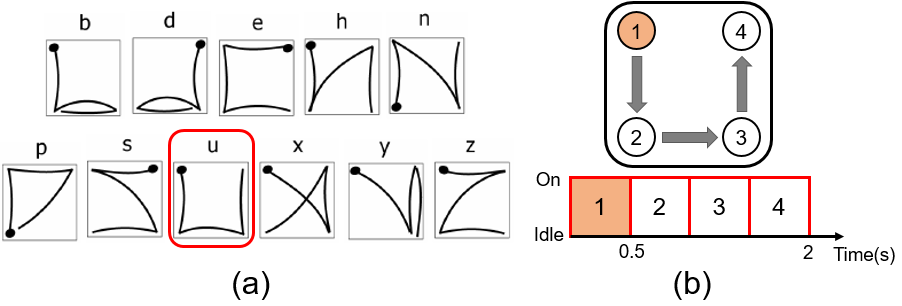}
  \caption{Pattern set used in Experiment 1. (a) 11 EdgeWrite alphabet patterns, (b) spatiotemporal description of the EdgeWrite pattern for letter U. The pattern starts from the colored point.}
  \label{Patternsetexp1}
  \Description[Tactile pattern set in the preliminary study]{The shape of eleven EdgeWrite alphabet patterns used in the recognition task of the preliminary study. The displaying timeline with no inter-stimulus interval is also shown.}
\end{figure}

From the 36 alphanumeric EdgeWrite \cite{wobbrock2003edgewrite} patterns used in Liao et al.'s \cite{liao2016edgevib} study, we selected 11 alphabet patterns with four vibration counts. The duration of each burst was 0.5 s, and there was no inter-stimulus interval (ISI). The spatiotemporal description of the pattern set is shown in Figure \ref{Patternsetexp1}.

\subsection{Participants}

We recruited 12 participants (2 females, mean age of 24.1 and SD of 4.49) from the university's public online community. All participants were right-handed. The participants were paid approximately \$20 for participating in this IRB-approved experiment.

\subsection{Procedure}

\begin{figure}[t]
  \centering
  \includegraphics[width=7cm]{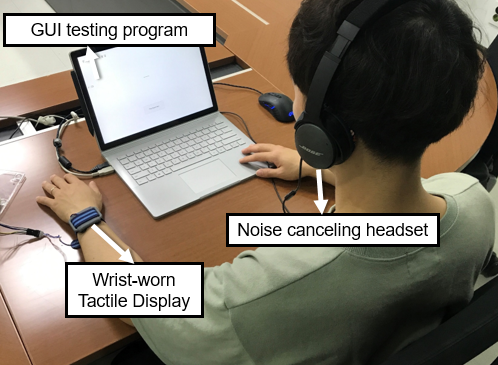}
  \caption{The experimental setup.}
  \label{Expsetup}
  \Description[Experimental setup]{A photo of the experimental environment. A person is wearing a noise-canceling headset with the wrist-worn tactile display prototype put on. He is sitting on a chair and watching the laptop on the desk to see a GUI testing program.}
\end{figure}

We guided the participants to a silent room and gave them the experimental guidelines. The participants wore the WTD prototype on their left wrist while sitting on a chair. The prototype was worn below the head of the ulna. They were encouraged to self-adjust the position or tightness of the strap to perceive the STPs as best as possible during the experiment and were asked to rest their arm comfortably as depicted in Figure \ref{Expsetup}. After being guided on how to handle the GUI experiment program, the participants wore noise-canceling headphones playing a pink noise to block out the sounds of the vibration.

Before the training and testing session, a 15-min learning session was conducted to allow the participants to become familiar with the 11 EdgeWrite patterns. A random alphabet was repeatedly prompted on the screen, and the participants answered the corresponding EdgeWrite pattern by clicking on a 2 x 2 grid of circular buttons with a mouse. During, and even after, the learning session, the participants could see the pattern set with the printed table. During the training and testing session, the participants were asked to answer the random pattern displayed on the WTD for every trial. They pressed the space key to play the pattern, submitted the answer with the alphabet keys, and finally confirmed by pressing the enter key. The pattern to be answered could be played only once during each trial. Before confirmation, the participants were allowed to modify their answer by pressing the backspace key. Only during the training session, the participants could manually play the patterns they wanted by pressing the alphabet keys before pressing the space key. Visual feedback was provided to indicate the correct answer. A short break was given for every 20 trials. Between conditions, the participants had to take off the device and rest at least 1 min to relieve the fatigue.

The training and testing session consisted of 33 trials (11 patterns $\times$ 3 reps) and 55 trials (11 patterns $\times$ 5 reps), respectively. The order of the patterns was randomized. It took approximately 2 h to complete the entire experiment.

\subsection{Design \& Analysis}

The experiment was a 3 $\times$ 2 within-subjects design with the following independent variables and levels:

\begin{itemize}
  \item Arm posture: \textit{Forward}, \textit{Right}, \textit{Down}
  \item Type of reference frame: \textit{RF 1}, \textit{RF 2}
\end{itemize}

We counterbalanced the order of conditions using a balanced Latin square. We collected 3960 answers (55 trials $\times$ 6 conditions $\times$ 12 participants) and calculated the pattern recognition accuracy (AC) and reaction time (RT) for each condition. The RT was measured as the time from the end of the pattern transmission to the moment the submission of the answer was confirmed by the enter key. For analysis, we performe a two-way repeated-measures ANOVA on the AC and RT. A pairwise t-test with a Bonferroni correction was used for a post hoc comparison.

\subsection{Results}

\begin{figure}[t]
  \centering
  \includegraphics[width=8cm]{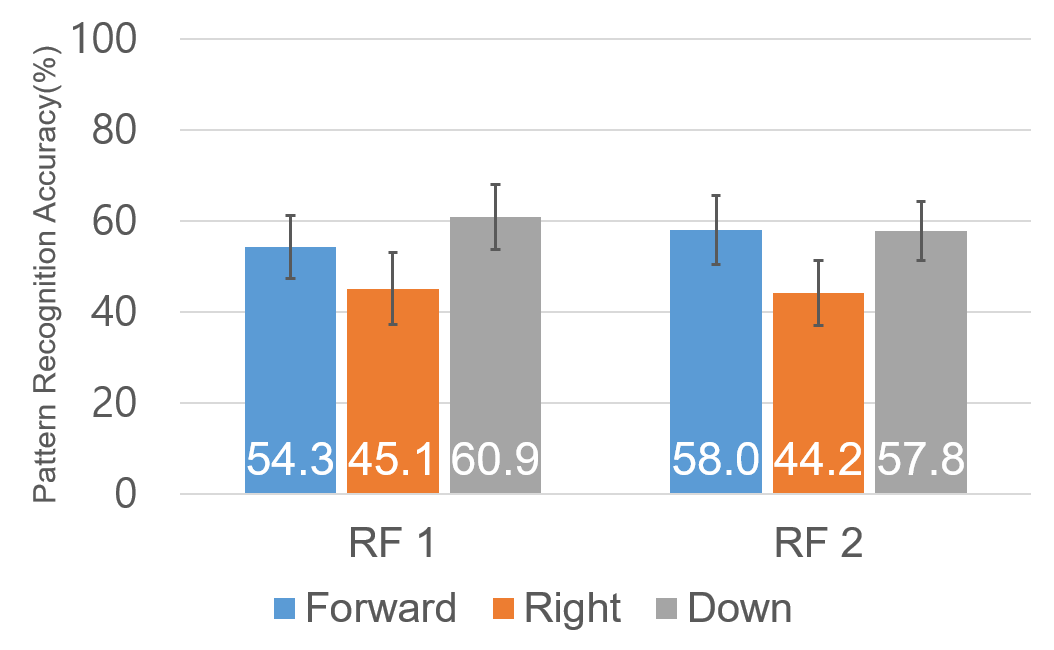}
  \caption{Mean of AC (\%) from the preliminary study. The error bars show standard errors.}
  \label{Resultexp1}
  \Description[The bar chart of accuracy results from the preliminary study]{The bar chart of the accuracies from the preliminary study. The accuracies for each condition is shown in Table 1.}
\end{figure}

\begin{table}[t]
  \setlength{\tabcolsep}{4pt} 
  \begin{tabular} {cccccccc}
    \hline\hline
    & & \textit{RF 1} & & & & \textit{RF 2} & \\
    \cline{2-4} \cline{6-8}
    & \textit{Forward} & \textit{Right} & \textit{Down} & & \textit{Forward} & \textit{Right} & \textit{Down} \\
    \hline
    AC & 54.3 & 45.1 & 60.9 & & 58.0 & 44.2 & 57.8   \\
    IT & 1.18 & 0.84 & 1.40 & & 1.37 & 0.91 & 1.39   \\
    RT & 3.4 & 3.3 & 3.1 & & 2.9 & 3.6 & 3.2   \\
    \hline\hline
    \end{tabular}
    \label{preliminaryResult}
    \caption{Mean of AC (\%), IT (bits), and RT (s) from the preliminary study.}
  \Description[The table for accuracy, information transfer, and reaction times from the preliminary study.]{The table contains information about accuracy, information transfer, and reaction time from the preliminary study.}
\end{table}
  
In the case of the AC
, the effect of the arm posture (\textit{F}(2,22) = 20.482, \textit{p} < .000) was significant. The post hoc comparison revealed that the AC of the \textit{Right} arm posture was significantly lower than that of \textit{Forward} (\textit{t} = 4.399, \textit{p} < .005) and \textit{Down} (\textit{t} = -5.431, \textit{p} < .005), when the ACs in \textit{RF 1} and \textit{RF 2} were averaged. The effect of the type of reference frame was not significant. There was no significant interaction effect. For the RT data, no independent variable showed a significant effect. The confusion matrices for each condition are attached in the Appendix section.

\subsection{Conclusion}

The results revealed that the arm posture can significantly affect the recognition accuracy of the STPs. This implies that the different arm postures need to be considered when designing and evaluating the STP pattern set. Based on this finding, we designed User Studies 1 and 2 to evaluate the effect of Heterogeneous Stroke considering different arm postures.

\section{User Study 1}

We designed User Study 1 to investigate the effect of Heterogeneous Stroke in terms of accuracy improvement. We set the arm posture and the Heterogeneous Stroke method as independent variables. Two arm postures, \textit{Forward} and \textit{Right}, which showed a significant difference in accuracy in the preliminary study were chosen. Three Heterogeneous Stroke methods, i.e., \textit{Baseline}, \textit{2-Hetero}, and \textit{4-Hetero}, depicted in Figure \ref{Threemethods}, were tested. The apparatus applied was identical to that of the preliminary study except that linear resonant actuators (LRAs) of 10 mm in diameter and with a 170 Hz resonance frequency (DMJBRN1036CB from Samsung Electro-mechanics) were used to control the vibration frequency.

\subsection{Tactile Pattern Set}

\begin{figure}[t]
  \centering
  \includegraphics[width=8cm]{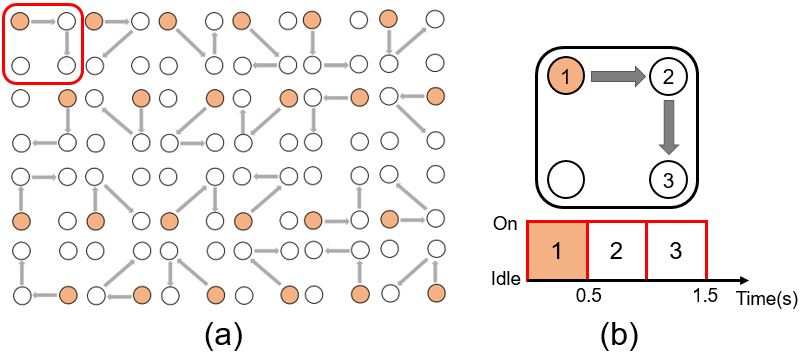}
  \caption{Three-point-stroke set used in User Study 1. (a) 24 three-point-stroke patterns, (b) spatiotemporal description of one pattern. The pattern starts from the colored point.}
  \label{Patternsetexp2}
  \Description[Tactile pattern set in the User Study 1]{The shape of 24 three point stroke patterns used in the recognition task of the User Study 1. The displaying timeline with no inter-stimulus interval is also shown.}
\end{figure}

We used the three-point-stroke set, which consists of all possible three consecutive independent points in a 2 $\times$ 2 tactor array so that all tactors can be used in a symmetrical manner. Because these patterns are basic elements of STPs with any 2D shape, we expect that it makes the empirical result of this study more generalizable. The spatiotemporal description of the pattern set is shown in Figure \ref{Patternsetexp2}.

\subsection{Participants}

We recruited 12 participants (2 females, mean age of 23.8 and SD of 3.64) from the university's public online community. One participant was left-handed but wore a watch on the left wrist. The participants were paid approximately \$50 for participating in this IRB-approved experiment.

\subsection{Procedure}

The procedure was similar to that of the preliminary study except that there was no learning session, and only the mouse was used to control the GUI experiment program. The participants clicked the play button to display the pattern, submitted their answer by consecutively clicking three points on the 2 $\times$ 2 grid of circular buttons, and finally confirmed by clicking the confirm button on the screen. Both the training and testing sessions consisted of two blocks of 48 trials (24 patterns $\times$ 2 reps). The study was conducted for 2 days, for approximately 2 h for each day.

\subsection{Design \& Analysis}

The experiment used a 3 $\times$ 2 within-subjects design with the following independent variables and levels:

\begin{itemize}
  \item Arm posture: \textit{Forward}, \textit{Right}
  \item Heterogeneous Stroke method: \textit{Baseline}, \textit{2-Hetero}, \textit{4-Hetero}
\end{itemize}

We counterbalanced the order of the conditions using a balanced Latin square. We collected 6912 answers (48 trials $\times$ 2 blocks $\times$ 6 conditions $\times$ 12 participants) and calculated the AC and RT. For the analysis, we performed a two-way repeated-measures ANOVA on the AC and RT. Because the AC violated the normality assumption, we applied an aligned rank transform (ART) before conducting the RM-ANOVA. For a post hoc comparison, a pairwise t-test or Wilcoxon signed-rank test with a Bonferroni correction was used depending on the result of the normality test.

\subsection{Results}

\begin{figure}[t]
  \centering
  \includegraphics[width=7.6cm]{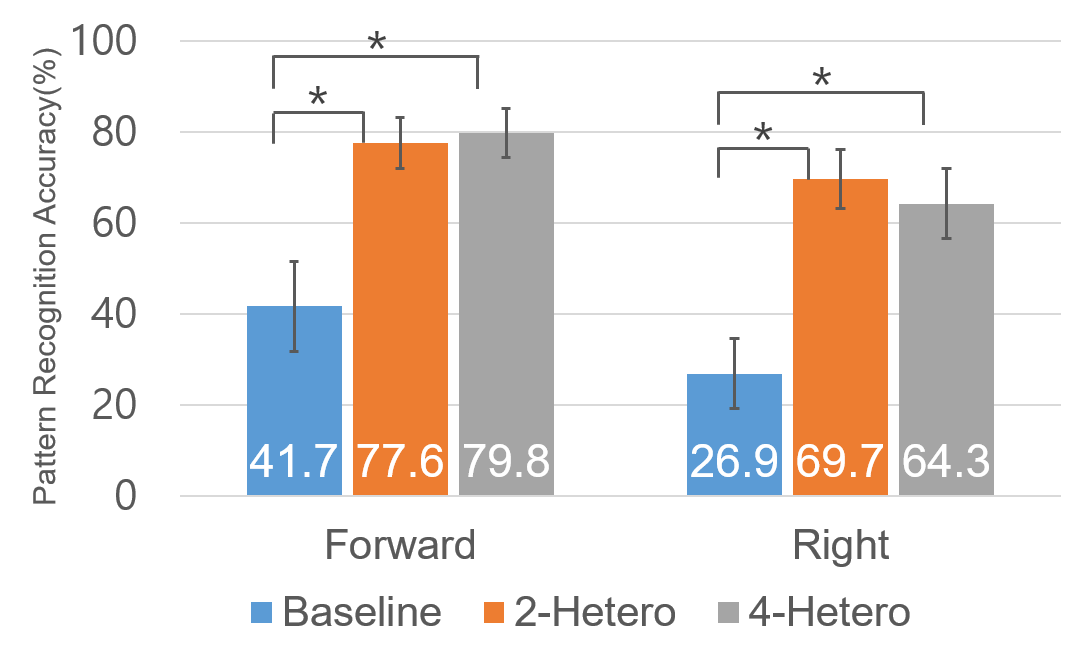}
  \caption{Mean of AC (\%) from the User Study 1. The error bars show standard errors, and the asterisks indicate significant differences (\textit{p} < .05).}
  \label{Resultexp2}
   \Description[The bar chart of accuracy results from the User Study 1.]{The bar chart of the accuracies from the User Study 1. The accuracies for each condition is shown in Table 2.}
\end{figure}

\begin{table}[t]
  \setlength{\tabcolsep}{2pt} 
  \begin{tabular} {cccccccc}
    \hline\hline
    & & \textit{Forward} & & & & \textit{Right} & \\
    \cline{2-4} \cline{6-8}
    & \textit{Baseline} & \textit{2-Hetero} & \textit{4-Hetero} & & \textit{Baseline} & \textit{2-Hetero} & \textit{4-Hetero} \\
    \hline
    AC & 41.7 & 77.6 & 79.8 & & 26.9 & 69.7 & 64.3     \\
    IT & 1.96 & 3.51 & 3.58 & & 1.59 & 3.23 & 2.97     \\
    RT & 3.1 & 2.9 & 3.2 & & 3.3 & 3.3 & 3.5     \\
    \hline\hline
    \end{tabular}
    \label{userStudy1Result}
    \caption{Mean of AC (\%), IT (bits), and RT (s) from User Study 1.}
    \Description[The table for accuracy, information transfer, and reaction times from the User Study 1.]{The table contains information about accuracy, information transfer, and reaction time from the User Study 1.}
\end{table}

In the case of the AC, the effect of the arm posture (\textit{F}(1,11) = 18.329, \textit{p} < .001) and the Heterogeneous Stroke method (\textit{F}(2,22) = 60.468, \textit{p} < .001) were significant. There was no significant interaction effect. A post hoc comparison revealed a significant difference between the \textit{Baseline} and other methods for both arm posture conditions: \textit{2-Hetero} (\textit{Z} = -3.061, \textit{p} < .05 for \textit{Forward} arm posture; \textit{Z} = -3.059, \textit{p} < .05 for \textit{Right}) and \textit{4-Hetero} (\textit{Z} = -3.059, \textit{p} < .05 for \textit{Forward}; and \textit{Z} = -3.061, \textit{p} < .05 for \textit{Right}). For the RT data, no independent variable showed a significant effect. The confusion matrices for each condition are attached in the Appendix section.

\subsection{Discussion}

The \textit{Baseline}, \textit{2-Hetero}, and \textit{4-Hetero} showed an accuracy of 34.3 \%, 73.7 \%, and 72.1 \%, respectively, when the ACs of both postures were averaged. Through our implementation of the Heterogeneous Stroke, the accuracy could be significantly improved, resulting in more than twice the value. In addition, the deviation in the accuracy was greatly reduced when comparing the accuracy ratio of the two postures in the \textit{Baseline} (0.65 = 26.9 \%/41.7 \%) and \textit{2-Hetero} (0.90 = 69.7 \%/77.6 \%). This implies that the overall accuracy improvement can reduce the inconsistency caused by the different postures.

The recognition task with the three-point-stroke set was challenging for which even a single confusion regarding the location of the tactor results in the wrong answer. Therefore, the overall accuracy was still low even when the \textit{2-Hetero} and \textit{4-Hetero} ($\sim$70 \%) were applied. In the following experiment, we apply Heterogeneous Stroke to the case of conveying alphanumeric characters with the EdgeWrite pattern set to examine if the proposed design can achieve a high performance with a complex STP set requiring an accurate shape recognition.

\section{User Study 2}

User Study 2 was designed to see whether the alphanumeric characters can be delivered with high accuracy by applying Heterogeneous Stroke. We chose the tactile EdgeWrite pattern set used in Liao et al.'s \cite{liao2016edgevib} study because it covers both the complete alphabets and numbers. We also followed the experimental design of the study including the pattern set and other details such as the vibration motor model (Precision Microdrive 310-113) to benefit from its formulated experimental settings.

We set the arm posture and the Heterogeneous Stroke method as independent variables. The two arm postures used in User Study 1 were tested again. Because there was no significant difference in accuracy between the \textit{2-Hetero} and \textit{4-Hetero} methods, only the \textit{Baseline} and \textit{2-Hetero} methods were tested.

\subsection{Participants}

We recruited 12 participants (5 females, mean age of 23.0 and SD of 4.05) for the alphabet group and 12 participants (5 females, mean age of 23.7 and SD of 2.87) for the digit group. In each group, one participant was left-handed but wore a watch on their left wrist. The participants in the alphabet and digit groups were paid approximately \$50 and \$20, respectively, for participating in the IRB-approved experiment.

\subsection{Tactile Pattern Set}

\begin{figure}[t]
  \centering
  \includegraphics[width=8.5cm]{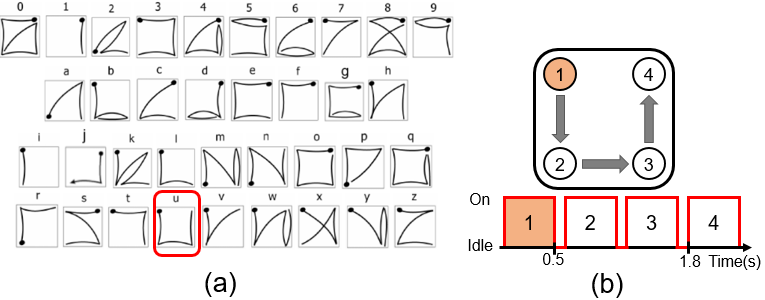}
  \caption{The pattern set used in User Study 2. (a) 26 EdgeWrite alphabet and 10 digit patterns, (b) spatiotemporal description of EdgeWrite pattern for letter U. The pattern starts from the colored point.}
  \label{Patternsetexp3}
  \Description[Tactile pattern set in the User Study 2]{The shape of 26 EdgeWrite alphabet and 10 EdgeWrite digit patterns used in the recognition task of the User Study 2. The display timeline with a 100 ms of inter-stimulus interval is also shown.}
\end{figure}

We used 26 EdgeWrite alphabet and 10 EdgeWrite digit patterns for each group. Unlike User Study 1, there was an ISI of 0.1 s between bursts to follow the experimental detail used in Liao et al.'s \cite{liao2016edgevib} study. The spatiotemporal description of the pattern set is shown in Figure \ref{Patternsetexp3}.

\subsection{Procedure}

Several details were modified from a preliminary study to reproduce the experimental settings of EdgeVib \cite{liao2016edgevib}. During the training session, the participants could repeatedly display the pattern by pressing the space key in each trial. The visual feedback for the correct answer was also provided during the testing session. For the alphabet group, the training and testing session consisted of 52 trials (26 patterns $\times$ 2 reps) and 104 trials (26 patterns $\times$ 4 reps). For the digit group, the training and testing session consisted of 20 trials (10 patterns $\times$ 2 reps) and 50 trials (10 patterns $\times$ 5 reps). The study for the alphabet group was conducted for two days (approximately 1.5 h each day), and the study for the digit group was conducted in a single day (approximately 2 h).

\subsection{Design \& Analysis}

The experiment was a 2 $\times$ 2 within-subjects design with following independent variables and levels: 

\begin{itemize}
  \item Arm posture: \textit{Forward}, \textit{Right}
  \item Heterogeneous Stroke method: \textit{Baseline}, \textit{2-Hetero}
\end{itemize}

We counterbalanced the order of the conditions using a balanced Latin square. We collected 4992 answers (104 trials $\times$ 4 conditions $\times$ 12 participants) for the alphabet group and 2400 answers (50 trials $\times$ 4 conditions $\times$ 12 participants) for the digit group. We calculated the AC and RT for each condition. We excluded one outlier subject in each group from the analysis who showed an AC outside 2 sigmas under at least one condition. We performed a two-way repeated-measures ANOVA on the AC and RT. Because the AC of the digit group violated the normality assumption, we performed an aligned rank transform (ART) before applying the RM-ANOVA. For a post hoc comparison, a pairwise t-test or Wilcoxon signed-rank test with a Bonferroni correction was used depending on the result of the normality test.

\subsection{Results}

The main results are summarized below and the confusion matrices for each group and condition are attached in the Appendix section.

\begin{figure}[t]
  \centering
  \includegraphics[width=8cm]{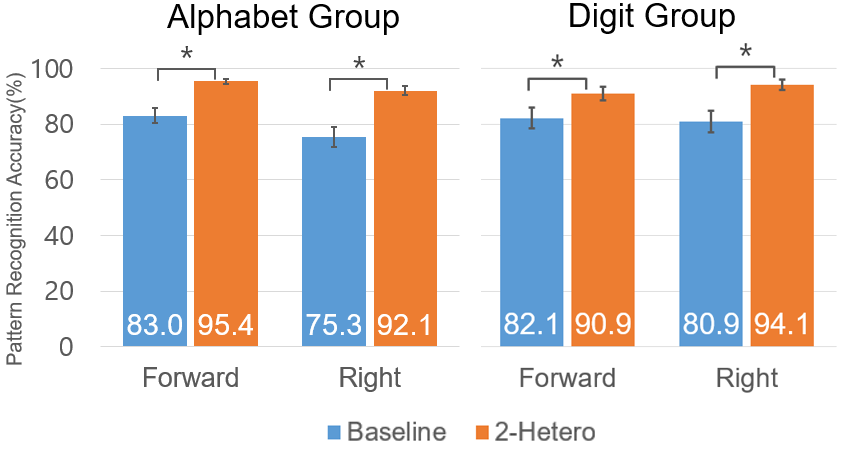}
  \caption{Mean of AC (\%) from the User Study 2. The error bars show standard errors, and the asterisks indicate significant differences (\textit{p} < .005).}
  \label{Resultexp3}
  \Description[The bar chart of accuracy results from the User Study 2.]{The bar chart of the accuracies from the User Study 2. The accuracies for each condition is shown in Table 3.}
\end{figure}

\begin{table}[t]
  \setlength{\tabcolsep}{6pt} 
  \begin{tabular} {cccccc}
    \hline\hline
    & \multicolumn{5}{c}{Alphabet Group} \\
    \cline{2-6}
    & \multicolumn{2}{c}{\textit{Forward}} & & \multicolumn{2}{c}{\textit{Right}} \\
    \cline{2-3} \cline{5-6}
    & \textit{Baseline} & \textit{2-Hetero} & & \textit{Baseline} & \textit{2-Hetero} \\
    \hline
    AC & 83.0 & 95.4 & & 75.3 & 92.1     \\
    IT & 3.71 & 4.42 & & 3.44 & 4.13     \\
    RT & 3.2 & 3.4 & & 3.5 & 3.4     \\
    \hline \\ 
    \hline\hline
    & \multicolumn{5}{c}{Digit Group} \\
    \cline{2-6}
    & \multicolumn{2}{c}{\textit{Forward}} & & \multicolumn{2}{c}{\textit{Right}} \\
    \cline{2-3} \cline{5-6}
    & \textit{Baseline} & \textit{2-Hetero} & & \textit{Baseline} & \textit{2-Hetero} \\
    \hline
    AC & 82.1 & 90.9 & & 80.9 & 94.1     \\
    IT & 2.41 & 2.67 & & 2.39 & 2.81     \\
    RT & 3.3 & 4.4 & & 2.9 & 4.0     \\
    \hline\hline
    \end{tabular}
    \label{userStudy2Result}
    \caption{Mean of AC (\%), IT (bits), and RT (s) from the User Study 2.}
    \Description[The table for accuracy, information transfer, and reaction times from the User Study 2.]{The table contains information about accuracy, information transfer, and reaction time from the User Study 2.}
\end{table}

\subsubsection{Alphabet Group}

In the case of the AC, the effect of the arm posture (\textit{F}(1,10) = 22.526, \textit{p} < .005) and the Heterogeneous Stroke method (\textit{F}(1,10) = 21.621, \textit{p} < .005) were significant. There was no significant interaction effect. The effect of the arm posture (\textit{F}(1,11) = 10.5488, \textit{p} < .005) and Heterogeneous Stroke method (\textit{F}(1,11) = 61.9274, \textit{p} < .001) was still significant even when the data of the outlier were included. For the RT data, no independent variable showed a significant effect.

\subsubsection{Digit Group}

In the case of the AC, the effect of the Heterogeneous Stroke method (\textit{F}(1,10) = 18.441, \textit{p} < .001) was significant. The effect of the arm posture was not significant. There was no significant interaction effect. The effect of the Heterogeneous Stroke method (\textit{F}(1,11) = 20.646, \textit{p} < .001) was still significant even when the data of the outlier were included. For the RT data, no independent variable showed a significant effect. 

\subsection{Discussion}

Using the \textit{2-Hetero} method, we achieved high accuracies of 93.8 \% and 92.4 \% on average in the alphabet and digit group, respectively.

Most of the apparatus and procedures were similar to those of the EdgeVib study \cite{liao2016edgevib}. The condition with the \textit{Right} arm posture and \textit{Baseline} is the equivalent condition of User Study 2 of the EdgeVib study. Comparing the results, we observed a slightly higher accuracy in our study (alphabets, 70.7 \% \cite{liao2016edgevib} versus 75.3 \%; and digits, 78.5 \% \cite{liao2016edgevib} versus 80.9 \%). Considering that the subjects were able to perform more training owing to the repeated measure design in our study, mostly consistent results were obtained. In the EdgeVib study, the duration of the EdgeWrite patterns was increased by inserting a delimiter in the middle of the pattern, forming a multi-stroke pattern to improve the accuracy. The alphabet and digit patterns with an average duration of 2.2 and 2.4 s were increased to 2.9 and 3.3 s, respectively, and the accuracy could be improved to 85.9 \% and 88.6 \% with these EdgeVib patterns. In comparison, the Heterogeneous Stroke design achieved a high accuracy (> 90\%) without lengthening the duration of the pattern.

\begin{table}[b]
  \caption{Number of confusions in frequency and roughness under \textit{4-Hetero} from User Study 1.}
  \label{tab:commands}
  \begin{tabular} {c|ccc|c}
    \hline\hline
         & Frequency & Roughness & Both & Total \\
    \hline
     Forward      & 228 & 46 & 14 & 288 \\
     Right       & 532 & 66 & 34 & 632 \\
    \hline
     Total       & 760 & 110 & 48 & 918 \\
    \hline\hline
    \end{tabular}
    \label{Confusions}
    \Description[Number of confusion from User Study 1.]{The table shows the number of confusion in frequency and roughness under 4-Hetero condition of User Study 2. Most of the confusion occurred by misinterpreting the frequency of vibration.}
\end{table}

\section{Discussion}

\subsection{Comparison between 2-Hetero and 4-Hetero methods}

The results of User Study 1 showed no significant difference in accuracy between the \textit{2-Hetero} and \textit{4-Hetero} methods. Although the number of unique vibrations was increased to further reduce the confusion, there was no additional improvement in accuracy with \textit{4-Hetero}. We asked the participants the following question after the study, \textit{"Which type of method made it easiest to recognize the pattern?"} Nine of the participants (75 \%) responded that the \textit{4-Hetero} method was the easiest, and three participants (25 \%) responded that the \textit{2-Hetero} method was the easiest. Whereas a majority of the participants preferred the \textit{4-Hetero} method, more than a few participants clearly preferred the \textit{2-Hetero} method.

To the survey question asking about the distinction of each vibrotactile parameter, all 12 participants responded that the roughness was well distinguished, whereas only 7 of the participants (58 \%) responded that the frequency was well distinguished. The rest responded with \textit{"I'm not sure"} (four participants, 33 \%) or it was \textit{"barely distinguishable"} (one participant, 8 \%). We analyzed the experimental results using a confusion matrix and found that 760 out of 918 moments of confusion (82.8 \%) occurred from misinterpreting the frequency (Table \ref{Confusions}). From the survey and confusion data, we observed that the levels of frequency were difficult to perceptually distinguish in comparison to the roughness. We expect this to be the main cause of no significant difference in accuracy between the \textit{2-Hetero} and \textit{4-Hetero} methods.

In the interview conducted after the experiment, the participants who preferred the \textit{4-Hetero} method generally agreed that the difference in frequency was relatively difficult to distinguish, but it was still helpful. P4 stated that \textit{"At the beginning of the experiment, two frequencies were almost indistinguishable, but as the experiment progressed, it slowly became possible to distinguish them. It was therefore easier to using 4-Hetero than 2-Hetero."} By contrast, those who preferred the \textit{2-Hetero} method pointed out that the recognition process became complicated. P6 stated that \textit{"With 4-Hetero, the addition of a difficult hint (frequency) was rather disturbing when I had to focus on the locus of the stimulus."} We expect that there is an advantage that can be obtained by increasing the number of ``hints'' and a disadvantage owing to the mental load incurred when perceiving and interpreting the stimulus simultaneously. If an added hint can make the stimuli sufficiently distinctive, the advantage will be greater than the disadvantage. Otherwise, the tradeoff between the two should be considered when designing Heterogeneous Stroke.

\subsection{Effect of posture on pattern recognition accuracy}

The arm posture of the subject showed a significant effect on the pattern recognition accuracy in the preliminary study and User Study 1, and in the alphabet group of User Study 2. Compared with other postures, the \textit{Right} posture showed a significantly lower accuracy. The \textit{Right} posture can be applied frequently in life when checking the time with a wristwatch. Previous WTD studies were conducted using different arm postures, including \textit{Forward} \cite{lee2015investigating, shim2018exploring, shim2019using, ion2015skin}, \textit{Right} \cite{liao2016edgevib}, and \textit{Down} \cite{lee2010buzzwear} arm postures. Lower accuracies might be obtained in the studies that use a \textit{Right} posture. This finding implies that WTD studies need to evaluate the pattern recognition accuracy in consideration of the different postures. This also implies that, when comparing the accuracy in previous studies, the posture needs to be considered as an independent variable.

Lawson et al. \cite{lawson2014you} and Scocchia et al. \cite{scocchia2009influence} reported a significant increase in the response time and error rate in haptic object identification tasks when the arm crosses the midline of the body while the head is not facing the object. Lawson et al. explained that proprioceptive and haptic inputs are remapped into the coordinate system of an external space and that this remapping is more difficult when the body is in an unusual position (e.g., hand crossing the body midline and the head facing away from the hand).  Although the tasks conducted in our study were different from those of Lawson et al.’s, the \textit{Right} posture is similar with the ``unusual'' posture in Lawson et al.'s study in that it also makes the hand cross the body midline. We might interpret our results on the \textit{Right} posture as being incurred from a failure of remapping the tactile pattern into the external space, as shown in Lawson et al.’s and Scocchia et al.’s studies.

\subsection{Actuators for Heterogeneous Stroke}

We overdrove LRA and ERM vibration motors with a voltage of 5 V throughout the experiment to implement the proposed concept. In particular, in User Study 2, because an LRA motor with a single resonance frequency was used to produce vibrations with two different frequencies, we used an overdriving voltage of 9 V to balance the intensity of the 300-Hz vibration with that of the 170-Hz vibration. However, overdriving the system can result in a safety problem. In our testing environment, the duration of the vibration was short and the pause until the next vibration was long (e.g., for the alphabet group in User Study 2, a motor was actuated for approximately 0.5 s per 2.23 s on average), and there was sufficient rest between blocks and trials. Although there were no safety problems owing to the controlled setup in our studies, unsafe circumstances (e.g., overheating) can occur when the motor is frequently actuated with an overdriving voltage.

Hwang et al. \cite{hwang2013real} used a dual mode actuator (DMA) \cite{hwang2013apparatus} capable of producing vibration composed of two primary frequencies to express diverse vibrotactile pitches for music playback. DMA has a structure that includes two built-in mass-spring elements with different resonance frequencies. If DMA can be utilized, excessive overdriving can possibly be avoided when implementing Heterogeneous Stroke using different vibration frequencies.

\subsection{Effect of reference frame in the preliminary study}

In the preliminary study, we set the type of reference frame as an independent variable. Because there exists a difference in the spatial acuity in the transverse and longitudinal axis of the wrist \cite{oakley2006determining}, we expected that it can affect the pattern recognition accuracy. However, there was no significant effect of the type of reference frame on the accuracy.

The difference in the average accuracy between the two reference frames was only 0.1 \%. However, the standard deviation (SD) of the accuracy across the entire patterns was 11 \%; for some patterns, the accuracy was higher with \textit{RF 1} or with \textit{RF 2}. For instance, for letter \textit{x}, a 17.2 \% higher accuracy was found in \textit{RF 1}, and for letter \textit{z}, a 16.7 \% higher accuracy was found in \textit{RF 2}. This may imply that the type of reference frame may still affect the recognition accuracy depending on the design of the pattern set.

\section{CONCLUSION}

In this study, we present Heterogeneous Stroke, a design concept that assigns unique vibrotactile stimuli to each tactor to effectively recognize STPs. We designed four unique vibrations for each tactor of a 2 $\times$ 2 grid type WTD by combining two levels of frequency and roughness of vibration to implement the proposed concept. We then experimentally showed the effects of Heterogeneous Stroke in terms of accuracy improvement. We achieved a high accuracy  (93.8 \% for alphabets and 92.4 \% for digits) when delivering 26 alphabets and 10 digits through our implementation of Heterogeneous Stroke. We also observed that the arm posture can significantly affect the recognition accuracy of STPs on the wrist. This implies that future WTD studies need to consider different postures when designing and evaluating the interaction with STPs. We hope that the proposed design concept and the empirical finding of our study can contribute to the design of effective tactile communications on a WTD.

\begin{acks}
This work has been supported by the Unmanned Swarm CPS Research Laboratory program of Defense Acquisition Program Administration and Agency for Defense Development.(UD190029ED)
\end{acks}

\bibliographystyle{ACM-Reference-Format}
\bibliography{main}

@inproceedings{lee2015investigating,
author = {Lee, Jaeyeon and Han, Jaehyun and Lee, Geehyuk},
title = {Investigating the Information Transfer Efficiency of a 3x3 Watch-Back Tactile Display},
year = {2015},
isbn = {9781450331456},
publisher = {Association for Computing Machinery},
address = {New York, NY, USA},
url = {https://doi.org/10.1145/2702123.2702530},
doi = {10.1145/2702123.2702530},
booktitle = {Proceedings of the 33rd Annual ACM Conference on Human Factors in Computing Systems},
pages = {1229–1232},
numpages = {4},
location = {Seoul, Republic of Korea},
series = {CHI '15}
}

@inproceedings{shim2019using,
author = {Shim, Youngbo Aram and Park, Keunwoo and Lee, Geehyuk},
title = {Using Poke Stimuli to Improve a 3x3 Watch-Back Tactile Display},
year = {2019},
isbn = {9781450368254},
publisher = {Association for Computing Machinery},
address = {New York, NY, USA},
url = {https://doi.org/10.1145/3338286.3340134},
doi = {10.1145/3338286.3340134},
booktitle = {Proceedings of the 21st International Conference on Human-Computer Interaction with Mobile Devices and Services},
articleno = {23},
numpages = {8},
location = {Taipei, Taiwan},
series = {MobileHCI '19}
}

@inproceedings{liao2016edgevib,
author = {Liao, Yi-Chi and Chen, Yi-Ling and Lo, Jo-Yu and Liang, Rong-Hao and Chan, Liwei and Chen, Bing-Yu},
title = {EdgeVib: Effective Alphanumeric Character Output Using a Wrist-Worn Tactile Display},
year = {2016},
isbn = {9781450341899},
publisher = {Association for Computing Machinery},
address = {New York, NY, USA},
url = {https://doi.org/10.1145/2984511.2984522},
doi = {10.1145/2984511.2984522},
booktitle = {Proceedings of the 29th Annual Symposium on User Interface Software and Technology},
pages = {595–601},
numpages = {7},
location = {Tokyo, Japan},
series = {UIST '16}
}

@inproceedings{lee2009mobile,
author = {Lee, Seungyon Claire and Starner, Thad},
title = {Mobile Gesture Interaction Using Wearable Tactile Displays},
year = {2009},
isbn = {9781605582474},
publisher = {Association for Computing Machinery},
address = {New York, NY, USA},
url = {https://doi.org/10.1145/1520340.1520499},
doi = {10.1145/1520340.1520499},
booktitle = {CHI '09 Extended Abstracts on Human Factors in Computing Systems},
pages = {3437–3442},
numpages = {6},
location = {Boston, MA, USA},
series = {CHI EA '09}
}

@inproceedings{ion2015skin,
author = {Ion, Alexandra and Wang, Edward Jay and Baudisch, Patrick},
title = {Skin Drag Displays: Dragging a Physical Tactor across the User's Skin Produces a Stronger Tactile Stimulus than Vibrotactile},
year = {2015},
isbn = {9781450331456},
publisher = {Association for Computing Machinery},
address = {New York, NY, USA},
url = {https://doi.org/10.1145/2702123.2702459},
doi = {10.1145/2702123.2702459},
booktitle = {Proceedings of the 33rd Annual ACM Conference on Human Factors in Computing Systems},
pages = {2501–2504},
numpages = {4},
location = {Seoul, Republic of Korea},
series = {CHI '15}
}

@inproceedings{chen2008tactor,
author = {Chen, Hsiang-Yu and Santos, Joseph and Graves, Matthew and Kim, Kwangtaek and Tan, Hong Z.},
title = {Tactor Localization at the Wrist},
year = {2008},
isbn = {9783540690566},
publisher = {Springer-Verlag},
address = {Berlin, Heidelberg},
url = {https://doi.org/10.1007/978-3-540-69057-3_25},
doi = {10.1007/978-3-540-69057-3_25},
booktitle = {Proceedings of the 6th International Conference on Haptics: Perception, Devices and Scenarios},
pages = {209–218},
numpages = {10},
location = {Madrid, Spain},
series = {EuroHaptics '08}
}

@inproceedings{shim2018exploring,
author = {Shim, Youngbo Aram and Lee, Jaeyeon and Lee, Geehyuk},
title = {Exploring Multimodal Watch-Back Tactile Display Using Wind and Vibration},
year = {2018},
isbn = {9781450356206},
publisher = {Association for Computing Machinery},
address = {New York, NY, USA},
url = {https://doi.org/10.1145/3173574.3173706},
doi = {10.1145/3173574.3173706},
booktitle = {Proceedings of the 2018 CHI Conference on Human Factors in Computing Systems},
pages = {1–12},
numpages = {12},
location = {Montreal QC, Canada},
series = {CHI '18}
}

@inproceedings{lee2010buzzwear,
author = {Lee, Seungyon "Claire" and Starner, Thad},
title = {BuzzWear: Alert Perception in Wearable Tactile Displays on the Wrist},
year = {2010},
isbn = {9781605589299},
publisher = {Association for Computing Machinery},
address = {New York, NY, USA},
url = {https://doi.org/10.1145/1753326.1753392},
doi = {10.1145/1753326.1753392},
booktitle = {Proceedings of the SIGCHI Conference on Human Factors in Computing Systems},
pages = {433–442},
numpages = {10},
location = {Atlanta, Georgia, USA},
series = {CHI '10}
}

@inproceedings{wobbrock2003edgewrite,
author = {Wobbrock, Jacob O. and Myers, Brad A. and Kembel, John A.},
title = {EdgeWrite: A Stylus-Based Text Entry Method Designed for High Accuracy and Stability of Motion},
year = {2003},
isbn = {1581136366},
publisher = {Association for Computing Machinery},
address = {New York, NY, USA},
url = {https://doi.org/10.1145/964696.964703},
doi = {10.1145/964696.964703},
booktitle = {Proceedings of the 16th Annual ACM Symposium on User Interface Software and Technology},
pages = {61–70},
numpages = {10},
location = {Vancouver, Canada},
series = {UIST '03}
}

@inproceedings{pakkanen2008perception,
author = {Pakkanen, Toni and Lylykangas, Jani and Raisamo, Jukka and Raisamo, Roope and Salminen, Katri and Rantala, Jussi and Surakka, Veikko},
title = {Perception of Low-Amplitude Haptic Stimuli When Biking},
year = {2008},
isbn = {9781605581989},
publisher = {Association for Computing Machinery},
address = {New York, NY, USA},
url = {https://doi.org/10.1145/1452392.1452449},
doi = {10.1145/1452392.1452449},
booktitle = {Proceedings of the 10th International Conference on Multimodal Interfaces},
pages = {281–284},
numpages = {4},
location = {Chania, Crete, Greece},
series = {ICMI '08}
}

@article{scocchia2009influence,
author = {Lisa Scocchia and Natale Stucchi and Jack M Loomis},
title ={The Influence of Facing Direction on the Haptic Identification of Two-Dimensional Raised Pictures},
journal = {Perception},
volume = {38},
number = {4},
pages = {606-612},
year = {2009},
doi = {10.1068/p5881},
note ={PMID: 19522327},
URL = { 
        https://doi.org/10.1068/p5881
},
eprint = { 
        https://doi.org/10.1068/p5881
}
}

@inproceedings{brewster2004tactons,
author = {Brewster, Stephen and Brown, Lorna M.},
title = {Tactons: Structured Tactile Messages for Non-Visual Information Display},
year = {2004},
publisher = {Australian Computer Society, Inc.},
address = {AUS},
booktitle = {Proceedings of the Fifth Conference on Australasian User Interface - Volume 28},
pages = {15–23},
numpages = {9},
location = {Dunedin, New Zealand},
series = {AUIC '04}
}

@inproceedings{matscheko2010tactor,
  title={Tactor placement in wrist worn wearables},
  publisher = {IEEE Computer Society},
  address = {USA},
  author={Matscheko, Michael and Ferscha, Alois and Riener, Andreas and Lehner, Manuel},
  booktitle={International Symposium on Wearable Computers (ISWC) 2010},
  pages={1--8},
  year={2010},
  organization={IEEE}
}

@INPROCEEDINGS{paneels2013what,  author={S. {Panëels} and M. {Anastassova} and S. {Strachan} and S. P. {Van} and S. {Sivacoumarane} and C. {Bolzmacher}},  booktitle={2013 World Haptics Conference (WHC)}, publisher = {IEEE Computer Society}, address = {USA},  title={What's around me? Multi-actuator haptic feedback on the wrist},   year={2013},  volume={},  number={},  pages={407-412},  doi={10.1109/WHC.2013.6548443}}

@InProceedings{tang2020design,
author="Tang, Jo-Hsi
and Raffa, Giuseppe
and Chan, Liwei",
editor="Rau, Pei-Luen Patrick",
title="Design of Vibrotactile Direction Feedbacks on Wrist for Three-Dimensional Spatial Guidance",
booktitle="Cross-Cultural Design. User Experience of Products, Services, and Intelligent Environments",
year="2020",
publisher="Springer International Publishing",
address="Cham",
pages="169--182",
isbn="978-3-030-49788-0"
}

@article{jones2009vibrotactile,
author = {Lynette A Jones and Jacquelyn Kunkel and Erin Piateski},
title ={Vibrotactile Pattern Recognition on the Arm and Back},
journal = {Perception},
volume = {38},
number = {1},
pages = {52-68},
year = {2009},
doi = {10.1068/p5914},
    note ={PMID: 19323136},

URL = {https://doi.org/10.1068/p5914},
eprint = {https://doi.org/10.1068/p5914}
}

@InProceedings{turcott2018efficient,
author="Turcott, Robert
and Chen, Jennifer
and Castillo, Pablo
and Knott, Brian
and Setiawan, Wahyudinata
and Briggs, Forrest
and Klumb, Keith
and Abnousi, Freddy
and Chakka, Prasad
and Lau, Frances
and Israr, Ali",
editor="Prattichizzo, Domenico
and Shinoda, Hiroyuki
and Tan, Hong Z.
and Ruffaldi, Emanuele
and Frisoli, Antonio",
title="Efficient Evaluation of Coding Strategies for Transcutaneous Language Communication",
booktitle="Haptics: Science, Technology, and Applications",
year="2018",
publisher="Springer International Publishing",
address="Cham",
pages="600--611",
isbn="978-3-319-93399-3"
}

@inproceedings{zhao2018coding,
author = {Zhao, Siyan and Israr, Ali and Lau, Frances and Abnousi, Freddy},
title = {Coding Tactile Symbols for Phonemic Communication},
year = {2018},
isbn = {9781450356206},
publisher = {Association for Computing Machinery},
address = {New York, NY, USA},
url = {https://doi.org/10.1145/3173574.3173966},
doi = {10.1145/3173574.3173966},
booktitle = {Proceedings of the 2018 CHI Conference on Human Factors in Computing Systems},
pages = {1–13},
numpages = {13},
location = {Montreal QC, Canada},
series = {CHI '18}
}

@inproceedings{chen2018effect,
author = {Chen, Qin and Perrault, Simon T. and Roy, Quentin and Wyse, Lonce},
title = {Effect of Temporality, Physical Activity and Cognitive Load on Spatiotemporal Vibrotactile Pattern Recognition},
year = {2018},
isbn = {9781450356169},
publisher = {Association for Computing Machinery},
address = {New York, NY, USA},
url = {https://doi.org/10.1145/3206505.3206511},
doi = {10.1145/3206505.3206511},
booktitle = {Proceedings of the 2018 International Conference on Advanced Visual Interfaces},
articleno = {25},
numpages = {9},
location = {Castiglione della Pescaia, Grosseto, Italy},
series = {AVI '18}
}

@article{cody2010tactile,
  title={Tactile spatial acuity is reduced by skin stretch at the human wrist},
  author={Cody, Frederick WJ and Idrees, Raheel and Spilioti, Diamantina X and Poliakoff, Ellen},
  journal={Neuroscience Letters},
  volume={484},
  number={1},
  pages={71--75},
  year={2010},
  publisher={Elsevier}
}

@article{medina2010maps,
title = "From maps to form to space: Touch and the body schema",
journal = "Neuropsychologia",
volume = "48",
number = "3",
pages = "645 - 654",
year = "2010",
note = "The Sense of Body",
issn = "0028-3932",
doi = "https://doi.org/10.1016/j.neuropsychologia.2009.08.017",
url = "http://www.sciencedirect.com/science/article/pii/S0028393209003364",
author = "Jared Medina and H. Branch Coslett"
}

@article{longo2015implicit,
author = {Longo, Matthew and Mancini, Flavia and Haggard, Patrick},
year = {2015},
month = {07},
pages = {77-87},
title = {Implicit body representations and tactile spatial remapping},
volume = {160},
journal = {Acta psychologica},
doi = {10.1016/j.actpsy.2015.07.002}
}

@article{mancini2011supramodal,
title = "A supramodal representation of the body surface",
journal = "Neuropsychologia",
volume = "49",
number = "5",
pages = "1194 - 1201",
year = "2011",
issn = "0028-3932",
doi = "https://doi.org/10.1016/j.neuropsychologia.2010.12.040",
url = "http://www.sciencedirect.com/science/article/pii/S0028393210005841",
author = "Flavia Mancini and Matthew R. Longo and Gian Domenico Iannetti and Patrick Haggard"
}

@article{ho2007head,
author = {Ho, Cristy and Spence, Charles},
year = {2007},
month = {06},
pages = {136-41},
title = {Head orientation biases tactile localization},
volume = {1144},
journal = {Brain research},
doi = {10.1016/j.brainres.2007.01.091}
}

@article{pritchett2011perceived,
author = {Pritchett, Lisa and Harris, Laurence},
year = {2011},
month = {05},
pages = {229-34},
title = {Perceived touch location is coded using a gaze signal},
volume = {213},
journal = {Experimental brain research. Experimentelle Hirnforschung. Expérimentation cérébrale},
doi = {10.1007/s00221-011-2713-0}
}

@article{lawson2014you,
  title={Where you look can influence haptic object recognition},
  author={R. Lawson and Amy Boylan and Lauren L. Edwards},
  journal={Attention, Perception, \& Psychophysics},
  year={2014},
  volume={76},
  pages={559-574}
}

@article{blum2019getting,
author = {Blum, Jeffrey R. and Fortin, Pascal E. and Al Taha, Feras and Alirezaee, Parisa and Demers, Marc and Weill-Duflos, Antoine and Cooperstock, Jeremy R.},
title = {Getting Your Hands Dirty Outside the Lab: A Practical Primer for Conducting Wearable Vibrotactile Haptics Research},
year = {2019},
issue_date = {July-Sept. 2019},
publisher = {IEEE Computer Society Press},
address = {Washington, DC, USA},
volume = {12},
number = {3},
issn = {1939-1412},
url = {https://doi.org/10.1109/TOH.2019.2930608},
doi = {10.1109/TOH.2019.2930608},
journal = {EEE Trans. Haptics},
month = jul,
pages = {232–246},
numpages = {15}
}

@article{park2018haptic,
author = {Park, Gunhyuk and Cha, Hojun and Choi, Seungmoon},
year = {2018},
month = {07},
pages = {1-1},
title = {Haptic Enchanters: Attachable and Detachable Vibrotactile Modules and Their Advantages},
volume = {PP},
journal = {IEEE Transactions on Haptics},
doi = {10.1109/TOH.2018.2859955}
}

@ARTICLE{tan2020methodology,  author={H. Z. {Tan} and S. {Choi} and F. W. Y. {Lau} and F. {Abnousi}},  journal={Proceedings of the IEEE},   title={Methodology for Maximizing Information Transmission of Haptic Devices: A Survey},   year={2020},  volume={108},  number={6},  pages={945-965},  doi={10.1109/JPROC.2020.2992561}}

@article{tan1999information,
author = {Tan, Hong and Durlach, N and Reed, Charlotte and Rabinowitz, W},
year = {1999},
month = {09},
pages = {993-1008},
title = {Information transmission with a multifinger tactual display},
volume = {61},
journal = {Perception \& psychophysics}
}

@inproceedings{brown2006multidimensional,
author = {Brown, Lorna M. and Brewster, Stephen A. and Purchase, Helen C.},
title = {Multidimensional Tactons for Non-Visual Information Presentation in Mobile Devices},
year = {2006},
isbn = {1595933905},
publisher = {Association for Computing Machinery},
address = {New York, NY, USA},
url = {https://doi.org/10.1145/1152215.1152265},
doi = {10.1145/1152215.1152265},
booktitle = {Proceedings of the 8th Conference on Human-Computer Interaction with Mobile Devices and Services},
pages = {231–238},
numpages = {8},
location = {Helsinki, Finland},
series = {MobileHCI '06}
}

@article{weinstein1968intensive,
  title={Intensive and extensive aspects of tactile sensitivity as a function of body part, sex and laterality},
  author={Weinstein, Sidney},
  journal={The skin senses},
  volume={1},  number={},
  pages = {195–222},
  year={1968},
  publisher={Springfield, III: Thomas}
}

@article{hollins2000individual,
author = {Hollins, Mark and Bensmaia, Sliman and Karlof, Kristie and Young, Forrest},
year = {2012},
month = {04},
pages = {1534-1544},
title = {Individual differences in perceptual space for tactile textures: Evidence from multidimensional scaling},
volume = {62},
journal = {Perception and Psychophysics},
doi = {10.3758/BF03212154}
}

@article{azadi2014evaluating,
author = {Azadi, Mojtaba and Jones, Lynette A.},
title = {Evaluating Vibrotactile Dimensions for the Design of Tactons},
year = {2014},
issue_date = {January 2014},
publisher = {IEEE Computer Society Press},
address = {Washington, DC, USA},
volume = {7},
number = {1},
issn = {1939-1412},
url = {https://doi.org/10.1109/TOH.2013.2296051},
doi = {10.1109/TOH.2013.2296051},
journal = {EEE Trans. Haptics},
month = jan,
pages = {14–23},
numpages = {10}
}

@inproceedings{brown2005first,
author = {Brown, Lorna M. and Brewster, Stephen A. and Purchase, Helen C.},
title = {A First Investigation into the Effectiveness of Tactons},
year = {2005},
isbn = {0769523102},
publisher = {IEEE Computer Society},
address = {USA},
url = {https://doi.org/10.1109/WHC.2005.6},
doi = {10.1109/WHC.2005.6},
booktitle = {Proceedings of the First Joint Eurohaptics Conference and Symposium on Haptic Interfaces for Virtual Environment and Teleoperator Systems},
pages = {167–176},
numpages = {10},
series = {WHC '05}
}

@INPROCEEDINGS{park2011perceptual,  author={G. {Park} and S. {Choi}},  booktitle={2011 IEEE World Haptics Conference},   title={Perceptual space of amplitude-modulated vibrotactile stimuli}, publisher = {IEEE Computer Society}, address = {USA},   year={2011},  volume={},  number={},  pages={59-64},  doi={10.1109/WHC.2011.5945462}}

@INPROCEEDINGS{oakley2006determining,
  author={I. {Oakley} and  {Yeongmi Kim} and  {Junhun Lee} and  {Jeha Ryu}},
  booktitle={2006 14th Symposium on Haptic Interfaces for Virtual Environment and Teleoperator Systems}, 
  title={Determining the Feasibility of Forearm Mounted Vibrotactile Displays}, 
  year={2006},
  volume={},
  number={},
  pages={27-34},
  doi={10.1109/HAPTIC.2006.1627079}
}

@article{post1994perception,
  title={Perception of vibrotactile stimuli during motor activity in human subjects},
  author={Post, LJ and Zompa, IC and Chapman, CE},
  journal={Experimental brain research},
  volume={100},
  number={1},
  pages={107--120},
  year={1994},
  publisher={Springer}
}

@inproceedings{luzhnica2019optimising,
author = {Luzhnica, Granit and Veas, Eduardo},
title = {Optimising Encoding for Vibrotactile Skin Reading},
year = {2019},
isbn = {9781450359702},
publisher = {Association for Computing Machinery},
address = {New York, NY, USA},
url = {https://doi.org/10.1145/3290605.3300465},
doi = {10.1145/3290605.3300465},
booktitle = {Proceedings of the 2019 CHI Conference on Human Factors in Computing Systems},
pages = {1–14},
numpages = {14},
location = {Glasgow, Scotland Uk},
series = {CHI '19}
}

@article{reed2018phonemic,
  title={A Phonemic-Based Tactile Display for Speech Communication},
  author={C. Reed and H. Tan and Zachary D. Perez and E. C. Wilson and Frederico M. Severgnini and Jaehong Jung and J. S. Martinez and Yang Jiao and A. Israr and F. Lau and Keith Klumb and Robert Turcott and F. Abnousi},
  journal={IEEE Transactions on Haptics},
  year={2019},
  volume={12},
  pages={2-17}
}

@ARTICLE{tan2020acquisition,  author={H. Z. {Tan} and C. M. {Reed} and Y. {Jiao} and Z. D. {Perez} and E. C. {Wilson} and J. {Jung} and J. S. {Martinez} and F. M. {Severgnini}},  journal={IEEE Transactions on Haptics},   title={Acquisition of 500 English Words through a TActile Phonemic Sleeve (TAPS)},   year={2020},  volume={13},  number={4},  pages={745-760},  doi={10.1109/TOH.2020.2973135}}

@article{weisenberger1986sensitivity,
  title={Sensitivity to amplitude-modulated vibrotactile signals.},
  author={J. Weisenberger},
  journal={The Journal of the Acoustical Society of America},
  year={1986},
  volume={80 6},
  pages={
          1707-15
        }
}

@ARTICLE{hwang2013real,  author={I. {Hwang} and H. {Lee} and S. {Choi}},  journal={IEEE Transactions on Haptics},   title={Real-Time Dual-Band Haptic Music Player for Mobile Devices},   year={2013},  volume={6},  number={3},  pages={340-351},  doi={10.1109/TOH.2013.7}}

@misc{hwang2013apparatus,
  title={Apparatus and method for generating vibration pattern},
  author={Hwang, Hyokune and Choi, Inho and Kim, Sunuk and Sa, Jaecheon and Joung, Munchae},
  year={2013},
  month=jun # "~18",
  publisher={Google Patents},
  note={US Patent 8,466,778}
}

\appendix
\section{appendix}

\begin{figure}[h]
  \centering
  \includegraphics[width=8.5cm]{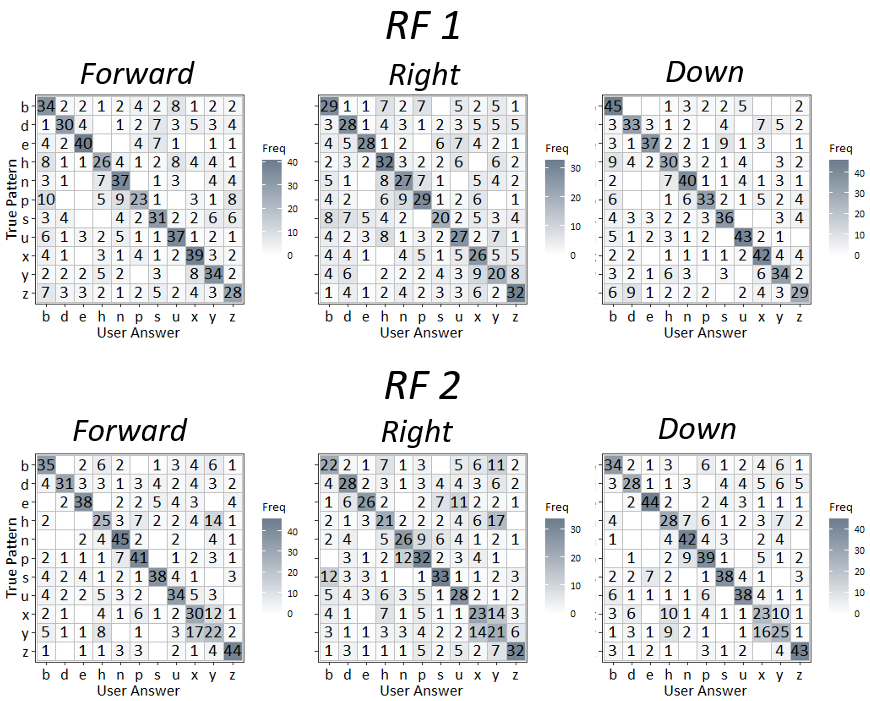}
  \caption{Stimulus-response confusion matrices of all condition from preliminary Study.}
  \label{ConfMatPreliminary}
\end{figure}

\begin{figure}
  \centering
  \includegraphics[width=8.5cm]{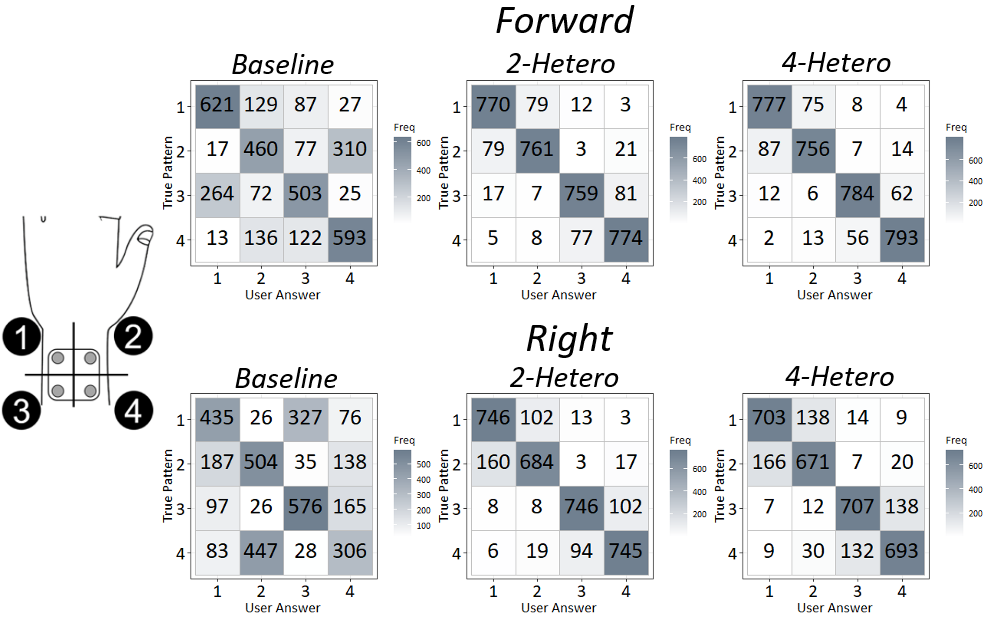}
  \caption{Stimulus-response confusion matrices of all condition from User Study 1. A three-point-stroke pattern displayed for each trial was counted as three stimuli to simplify the confusion matrix.}
  \label{ConfMatStudy1}
\end{figure}

\begin{figure}
  \centering
  \includegraphics[width=8.5cm]{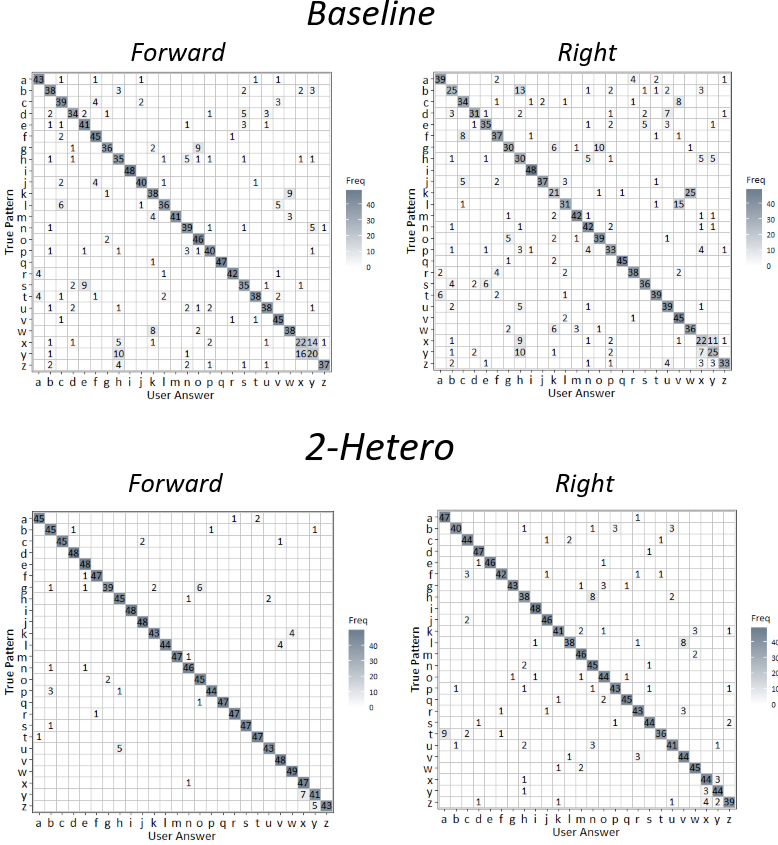}
  \caption{Stimulus-response confusion matrices of all condition of alphabet group from User Study 2.}
  \label{ConfMatStudy2Alphabet}
\end{figure}

\begin{figure}
  \centering
  \includegraphics[width=8.5cm]{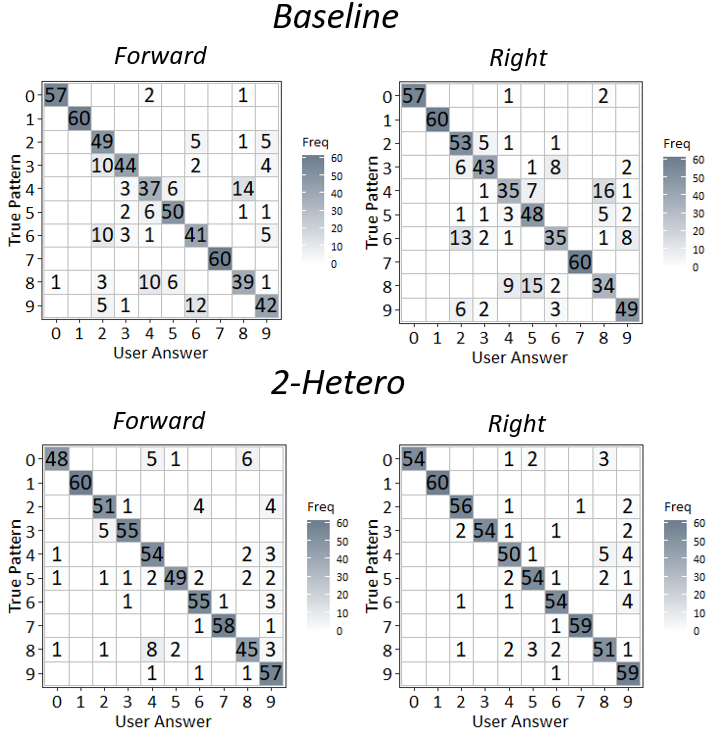}
  \caption{Stimulus-response confusion matrices of all condition of digit group from User Study 2.}
  \label{ConfMatStudy2Digit}
\end{figure}

Although omitted from the main text owing to a paper length issue, we attach confusion matrices here to help a clearer understanding of the experimental results of tactile pattern recognition tasks. Confusion matrices from the preliminary study and User Study 1 are shown in Figure \ref{ConfMatPreliminary} and \ref{ConfMatStudy1}, respectively. Confusion matrices from the User Study 2 are shown in Figure \ref{ConfMatStudy2Alphabet} and \ref{ConfMatStudy2Digit}.

\clearpage

\end{document}